\documentclass[a4paper, 11pt]{article}
\usepackage{pos}

\usepackage{graphicx}
\usepackage{subcaption}
\usepackage{braket}
\usepackage{placeins}
\usepackage{slashed}
\usepackage{wrapfig}
\usepackage[font=small, labelfont=bf, textfont={small,it}]{caption}
\newcommand{\beq}{\begin{eqnarray}}
\newcommand{\eeq}{\end{eqnarray}}
\newcommand{\beqnn}{\begin{eqnarray*}}
\newcommand{\eeqnn}{\end{eqnarray*}}

\newcommand{\Pf}{\mathrm{Pf}}
\newcommand{\SU}{\mathrm{SU}}

\newcommand{\MS}{{\scriptscriptstyle{\overline{\mathrm{MS}}}}}

\renewcommand{\P}{{\scriptscriptstyle{\mathrm{P}}}}
\renewcommand{\S}{{\scriptscriptstyle{\mathrm{S}}}}
\renewcommand{\t}{{\scriptscriptstyle{\mathrm{t}}}}
\newcommand{\A}{{\scriptscriptstyle{\mathrm{A}}}}
\newcommand{\R}{{\scriptscriptstyle{\mathrm{R}}}}
\newcommand{\RGI}{{\scriptscriptstyle{\mathrm{RGI}}}}
\newcommand{\W}{{\scriptscriptstyle{\mathrm{W}}}}
\newcommand{\PCAC}{{\scriptscriptstyle{\mathrm{PCAC}}}}
\newcommand{\NSVZ}{{\scriptscriptstyle{\mathrm{NSVZ}}}}

\newcommand{\s}{{\scriptscriptstyle{\mathrm{s}}}}
\newcommand{\I}{{\scriptscriptstyle{\mathrm{I}}}}
\newcommand{\E}{{\scriptscriptstyle{\mathrm{E}}}}
\newcommand{\Ep}{{\scriptscriptstyle{\mathrm{E}^\prime}}}
\newcommand{\zero}{{\scriptscriptstyle{(0)}}}
\newcommand{\m}{{\scriptscriptstyle{\rm m}}}
\newcommand{\cond}{\langle{\rm Tr}\lambda^2\rangle}

\title{The gluino condensate of large-$N$ SUSY Yang--Mills}
\ShortTitle{The gluino condensate of large-$N$ SUSY Yang--Mills}

\author*[a]{Claudio Bonanno}
\author[a,b]{Pietro Butti}
\author[a]{Margarita Garc\'ia P\'erez}
\author[a,c]{\\Antonio Gonz\'alez-Arroyo}
\author[d,e]{Ken-Ichi Ishikawa}
\author[e]{Masanori Okawa}

\affiliation[a]{Instituto de F\'isica T\'eorica UAM-CSIC, Calle Nicol\'as Cabrera 13-15,\\Universidad Aut\'onoma de Madrid, Cantoblanco, E-28049 Madrid, Spain}

\affiliation[b]{Departamento de F\'isica T\'eorica, Facultad de Ciencias and \\ Centro de Astropart\'iculas y F\'isica de Altas Energ\'ias (CAPA),  
	\\ Universidad de Zaragoza, Calle Pedro Cerbuna 12, E-50009, Zaragoza, Spain}

\affiliation[c]{Departamento de F\'isica Te\'orica, Universidad Aut\'onoma de Madrid,\\M\'odulo 15, Cantoblanco, E-28049 Madrid, Spain}

\affiliation[d]{Core of Research for the Energetic Universe,\\Graduate School of Advanced Science and Engineering,\\Hiroshima University, Higashi-Hiroshima, Hiroshima 739-8526, Japan}

\affiliation[e]{Graduate School of Advanced Science and Engineering, Hiroshima University,\\Higashi-Hiroshima, Hiroshima 739-8526, Japan}

\emailAdd{claudio.bonanno@csic.es}
\emailAdd{pbutti@unizar.es}
\emailAdd{margarita.garcia@csic.es}
\emailAdd{antonio.gonzalez-arroyo@uam.es}
\emailAdd{ishikawa@theo.phys.sci.hiroshima-u.ac.jp}
\emailAdd{okawa@hiroshima-u.ac.jp}

\abstract{We present the first lattice determination of the SUSY $\SU(N)$ Yang--Mills gluino condensate at large $N$. We exploit large-$N$ twisted volume reduction, and present two determinations based on the Banks--Casher relation and on a Gell-Mann--Oakes--Renner-like formula, both giving perfectly compatible results. By expressing the lattice results in the Novikov--Shifman--Vainshtein--Zakharov scheme, we are able for the first time to compare lattice and analytical computations, resolving a 40-year-long debate about the actual value and $N$-dependence of the gluino condensate.}

\FullConference{The 41$^{\text{st}}$ International Symposium on Lattice Field Theory (LATTICE2024)\\
28 July -- 3 August 2024\\
Liverpool, UK\\}

\begin{document}
\maketitle

\section{Introduction}

The $\mathcal{N}=1$ supersymmetric (SUSY) $\SU(N)$ Yang--Mills theory consists of a bosonic particle, the gluon, and its fermionic super-partner, the gluino. These particles are described in terms of a $\SU(N)$ gauge field coupled to 1 massless adjoint Majorana field. The SUSY Yang--Mills (SYM) action enjoys a global $\mathrm{U}(1)_{\scriptscriptstyle{\rm V}} \otimes \mathrm{U}(1)_{\scriptscriptstyle{\rm A}}$ flavor chiral symmetry. At the quantum level, the vector symmetry is exactly realized à la Wigner--Weyl, while the axial sub-group is anomalously and spontaneously broken with the following pattern:
\beq
\mathrm{U}(1)_{\scriptscriptstyle{\rm A}} \qquad \underset{\text{ Anomaly}}{\longrightarrow} \qquad \mathbb{Z}_{2N} \qquad \underset{\text{Spontaneous}}{\longrightarrow} \qquad \mathbb{Z}_{2}.
\eeq
The theory is thus expected to be characterized by a non-vanishing gluino condensate $\cond$ due to the spontaneous breaking of the residual discrete $\mathbb{Z}_{2N}$ symmetry.

The gluino condensate can be calculated exactly using analytic methods~\cite{Novikov:1983ee,Rossi:1983bu,Amati:1984uz,Novikov:1985ic}. However, its actual value is the subject of a debate~\cite{Hollowood:1999qn}. Two instanton-based calculations performed, respectively, in the Strong-Coupling (SC)~\cite{Novikov:1983ee,Rossi:1983bu,Amati:1984uz} and in the Weak-Coupling (WC)~\cite{Novikov:1985ic} regimes yielded the following for the Renormalization-Group-Invariant (RGI) condensate:
\beq\label{eq:Sigma_Inst}
\Sigma_\RGI\equiv \frac{1}{(4\pi) ^2 b_0 N} \left\vert \cond \right\vert= 
\begin{cases}  
2 \mathrm{e}  \, \Lambda^3_\NSVZ / N \qquad &\text{(SC)} \,,\\
\Lambda^3_\NSVZ \qquad &\text{(WC)} \,.
\end{cases}
\eeq
Here $\Lambda_{\NSVZ}$ is the dynamically-generated scale of SYM computed in the Novikov--Shifman--Vainshtein--Zakharov (NSVZ) scheme~\cite{Novikov:1983ee,Shifman:1986zi}. Adopting the standard QCD convention (see~\cite{Armoni:2003yv} to trivially map it to the SUSY one), it reads: 
\beq\label{eq:NSVZ_Lambda}
\Lambda_\NSVZ^3  = \frac{\mu^3}{b_0 \lambda_{\t}^{(\NSVZ)}(\mu) } \exp \left ( \frac {-8 \pi^2}{\lambda_{\t}^{(\NSVZ)}(\mu)} \right)\,,
\eeq
with $\lambda_{\t}^{(\NSVZ)}(\mu)$ the renormalized 't Hooft coupling in the NSVZ scheme, and $b_0=3/(4\pi)^2$ the first universal coefficient of the SYM $\beta$-function. Recently, an alternative calculation of the gluino condensate, based on the use of fractional instantons~\cite{tHooft:1981nnx,GarciaPerez:2000aiw,Gonzalez-Arroyo:2019wpu} has been carried out in~\cite{Anber:2022qsz}, based on ideas put forward in~\cite{Cohen:1983fd,Zhitnitsky:1989ds,Davies:1999uw,Davies:2000nw}. The authors found $\Sigma_\RGI= 2\Lambda_\NSVZ^3$ for gauge group $\SU(2)$ and argue that $2 \to N$ for $\SU(N)$, thus yielding yet another result.

This proceeding reports on the main results of our paper~\cite{Bonanno:2024bqg}, where we performed the first non-perturbative first-principles calculation of the gluino condensate of large-$N$ SYM theory via numerical Monte Carlo simulations of the lattice-discretized theory. Despite impressive recent progress in lattice simulations of SUSY theories~\cite{,Ali:2018fbq,Ali:2018dnd,Ali:2019agk,Bergner:2019dim,Piemonte:2020wkm,Bergner:2022snd,Schaich:2022xgy}, the lattice literature on this topic is quite limited~\cite{Giedt:2008xm,Kim:2011fw,Bergner:2019dim,Piemonte:2020wkm}, and our paper presents the first comparison between numerical and analytical results.

We here anticipate that our value and $N$-dependence for the gluino condensate agree with the WC prediction. After the publication of our paper~\cite{Bonanno:2024bqg}, the new study~\cite{Anber:2024mco} from the same authors of~\cite{Anber:2022qsz} appeared, reporting a value and $N$-dependence for $\Sigma_\RGI/\Lambda_{\NSVZ}^3$ in agreement with ours.

\section{From the lattice to the NSVZ scheme}

To compare numerical and analytical results, we must compute two quantities: the RGI gluino condensate $\Sigma_\RGI$ and the SUSY scale $\Lambda_\NSVZ$. We review their definition in the following.

\subsection{The dynamically-generated scale}
The $\Lambda$-parameter is a scheme-dependent quantity defined as the integration constant of the Callan--Symanzik equation for the renormalized 't Hooft coupling $\lambda_\t^{(\s)}$, expressed via the $\beta$-function:
\beq
\beta_\s\left(\lambda_\t^{(\s)}\right) = \frac{\mathrm{d} \, \lambda_\t^{(\s)}(\mu)}{\mathrm{d} \log(\mu^2)},
\eeq
\beq
\Lambda_\s = \mu\left[b_0
\lambda_\t^{(\s)}(\mu)\right]^\frac{-b_1}{2{b_0}^2}\exp\left(\frac{-1}{2b_0\lambda_\t^{(\s)}(\mu)}\right) 
 \exp \left[-\int_0^{\lambda_\t^{(\s)}(\mu)} \mathrm{d}x
\left(\frac{1}{2 \beta_s(x)}+\frac{1}{2 b_0 x^2}-\frac{b_1}{2b_0^2x}\right)\right]\,,
\eeq
with $b_0 = 3/(4\pi)^2$, $b_1 = 6/(4\pi)^4$ the first two universal (i.e., scheme-independent) coefficients in the perturbative expansion of the $\beta$-function. In the NSVZ scheme, the exact $\beta$-function
\beq
\beta_\NSVZ\left(\lambda_\t^{(\NSVZ)}\right) = - \frac{b_0 {\lambda_\t^{(\NSVZ)}}^2}{1-\frac{b_1}{b_0}\lambda_\t^{(\NSVZ)}}
\eeq
yields the exact expression for $\Lambda_\NSVZ$ in Eq.~\eqref{eq:NSVZ_Lambda}.

\subsection{The RGI gluino condensate}

The Callan--Symanzik equation for the renormalized gluino mass $m_\R^{(\s)}$, expressed via the anomalous dimension $\tau$-function, also features a \emph{scheme-independent} integration constant:
\beq
\tau_{\s}\left(\lambda_\t^{(\s)}\right) = \frac{{\rm d} \log\left(m_{\R}^{(\s)}(\mu)\right)}{{\rm d} \log(\mu)}\,.
\eeq
\beq
m_\RGI = \widetilde{{\cal A}} \, m_\R^{(\s)}(\mu) \, 
\left [2 b_0\lambda_\t^{(\s)}(\mu)\right]^{-\frac{d_0}{2b_0}} \exp \left[-\int_0^{\lambda_\t^{(\s)}(\mu)} \mathrm{d}x
\left(\frac{\tau_\s(x)}{2 \beta_\s(x)}-\frac{1}{ x}\right)\right]\,,
\eeq
with $d_0 = 2 b_0$ the first universal (i.e., scheme-independent) coefficient of the perturbative expansion of the $\tau$-function, and $\widetilde{{\cal A}}$ an arbitrary numerical constant. In the NSVZ scheme, also the $\tau$-function is known exactly~\cite{Jack:1997pa}:
\beq
\frac{\tau_\NSVZ(x)}{2 \beta_\NSVZ(x)} = \frac{1}{x (1 - b_1 x/ b_0) } \,.
\eeq
Since the product $\Sigma_\R^{(\s)}(\mu) m_\R^{(\s)}(\mu)$ is RGI---with $\Sigma_\R^{(\s)}(\mu)$ the usual scheme-/scale-dependent fermion condensate---the RGI gluino condensate can be easily defined as~\cite{Giusti:1998wy,DellaMorte:2005kg}:
\beq\label{eq:def_Sigma_RGI}
\Sigma_\RGI = {\cal A} \, \Sigma_\R^{(\s)}(\mu) \, 
\left [2 b_0\lambda_\t^{(\s)}(\mu)\right]^{\frac{d_0}{2b_0}} \exp \left[\int_0^{\lambda_\t^{(\s)}(\mu)} \mathrm{d}x
\left(\frac{\tau_\s(x)}{2 \beta_\s(x)}-\frac{1}{x}\right)\right]\,.
\eeq
Once the exact $\beta$ and $\tau$ functions of the NSVZ scheme are inserted in Eq.~\eqref{eq:def_Sigma_RGI}, the arbitrary constant $\mathcal{A}$ must be chosen as $\mathcal{A} = \frac{8\pi^2}{9 N^2}$ in order for Eq.~\eqref{eq:def_Sigma_RGI} to reproduce the known exact expression for the RGI condensate in terms of $\Sigma_{\R}^{(\NSVZ)}(\mu)$~\cite{Shifman:1986zi,Hisano:1997ua}:
\beq
\Sigma_{\RGI} = \frac{1}{(4\pi)^2 b_0 N} \left\vert\cond\right\vert, \qquad
\left\vert\cond\right\vert = \frac{\lambda_\t^{(\NSVZ)}(\mu)}{N\left[1-\lambda_\t^{(\NSVZ)} (\mu)/(8\pi^2)\right] } \Sigma_\R^{(\NSVZ)}(\mu).
\eeq

\section{Lattice setup}

Our large-$N$ calculation exploits large-$N$ twisted volume reduction. By virtue of the well-known dynamical equivalence between space-time and color degrees of freedom unvailed in the pioneering paper of Eguchi and Kawai~\cite{Eguchi:1982nm}, it is possible to simulate 4$d$ large-$N$ gauge theories as a matrix model which can be interpreted as standard lattice gauge theory defined on a reduced one-point space-time lattice with twisted boundary conditions~\cite{tHooft:1979rtg,Gonzalez-Arroyo:1982hyq,Gonzalez-Arroyo:2010omx}.

Reduced models have been extensively used in the last decade to study several large-$N$ gauge theories~\cite{Gonzalez-Arroyo:2012euf,GarciaPerez:2014azn,Perez:2020vbn,Bonanno:2023ypf}, including also theories with dynamical adjoint fermions~\cite{Gonzalez-Arroyo:2013bta,GarciaPerez:2015rda}. This enabled in Ref.~\cite{Butti:2022sgy} the study of large-$N$ SYM on the lattice using twisted volume reduction and the lattice techniques developed by the DESY--Jena--Regensburg--M\"unster collaboration, see Refs.~\cite{Ali:2019agk,Ali:2018fbq,Ali:2018dnd}. More precisley, in Ref.~\cite{Butti:2022sgy} we generated several gauge configurations for a few values of inverse bare 't Hooft coupling $b=1/\lambda_{{\scriptscriptstyle{\rm L}}}$ and $N$ according the following setup.
\vspace*{-0.1\baselineskip}
\begin{itemize}
\item Simulations are performed using a dynamical massive gluino, discretized using Wilson fermions. The gluino mass is controlled via the Wilson hopping parameter $\kappa$.
\item The mild sign problem introduced by the non-positivity of the Pfaffian of the lattice Wilson--Dirac operator is bypassed via sign-quenched simulations. Since no occurrence of negative signs of $\Pf(CD_\W)$ was observed in~\cite{Butti:2022sgy}, no reweighting was needed.
\item The lattice regularization and the non-zero gluino mass explicitly break SUSY. According to the Kaplan--Curci--Veneziano prescription, the SUSY-restoration limit is achieved as the joint continuum and chiral (massless gluino) limit~\cite{Kaplan:1983sk,Curci:1986sm}.
\item The massless gluino limit is obtained requiring a vanishing mass for the ``adjoint pion''. This is an unphysical particle which is introduced by supplementing SYM with a quenched valence gluino. This approach can be rigorously justified within the theoretical framework of Partially Quenched Chiral Perturbation Theory~\cite{Munster:2014cja}.
\end{itemize}

The gluino condensate is computed adopting two different methods:
\begin{itemize}
\item Banks--Casher (BC) formula~\cite{Leutwyler:1992yt}.
\beq
\frac{\Sigma_\R^{(\s)}(\mu)}{2\pi} = \lim_{\lambda\to 0}\lim_{m_\R\to0}\lim_{V\to\infty} \left[\rho_\R^{(\s)}(\mu)\right](\lambda_\R,m_\R) \, .
\eeq
Here $\rho_\R$ stands for the spectral density of eigenmodes $\mathrm{i}\lambda_\R+m_\R$ of the massive Dirac operator.
\item Gell-Mann--Oakes--Renner (GMOR) relation~\cite{Munster:2014cja}.
\beq
m_\pi^2 = 2\frac{\Sigma_\R^{(\s)}(\mu)}{F^2_\pi} m_\R^{(\s)}(\mu)\, .
\eeq
This relation involves the non-singlet adjoint pion mass $m_\pi$ and its decay constant $F_\pi$.
\end{itemize}

\section{Results}

In this section we summarize the main results of Ref.~\cite{Bonanno:2024bqg}, obtained for $b=0.340, 0.345, 0.350$ and $N=169, 289, 361$, using the gauge ensembles generated in Ref.~\cite{Butti:2022sgy}. Scale setting was performed in~\cite{Butti:2022sgy} using gradient flow through the standard reference scale $\sqrt{8 t_0}$.

\subsection{The gluino condensate from the BC relation}

\begin{figure}[!htb]
\begin{subfigure}{0.48\columnwidth}
\centering
\includegraphics[width=0.95\textwidth]{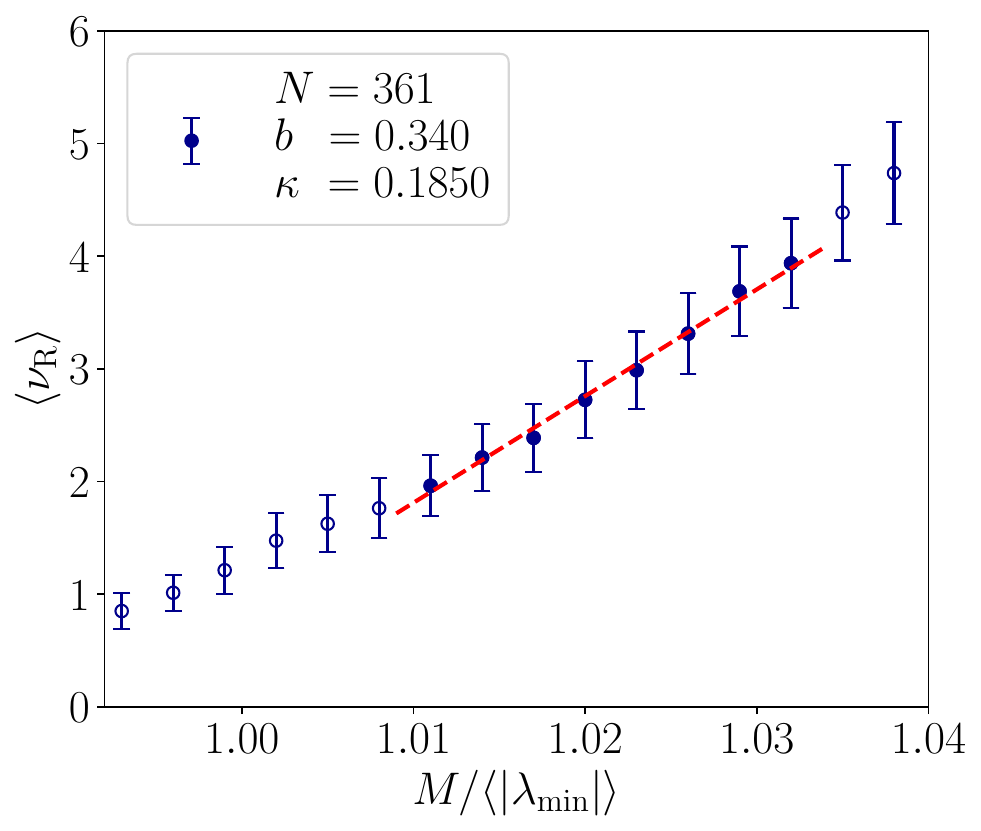}
\caption{Example of the linear fit to the mode number to extract the slope.}
\label{fig:fit_mode_number}
\end{subfigure}
\hfill
\begin{subfigure}{0.48\columnwidth}
\centering
\includegraphics[width=0.95\textwidth]{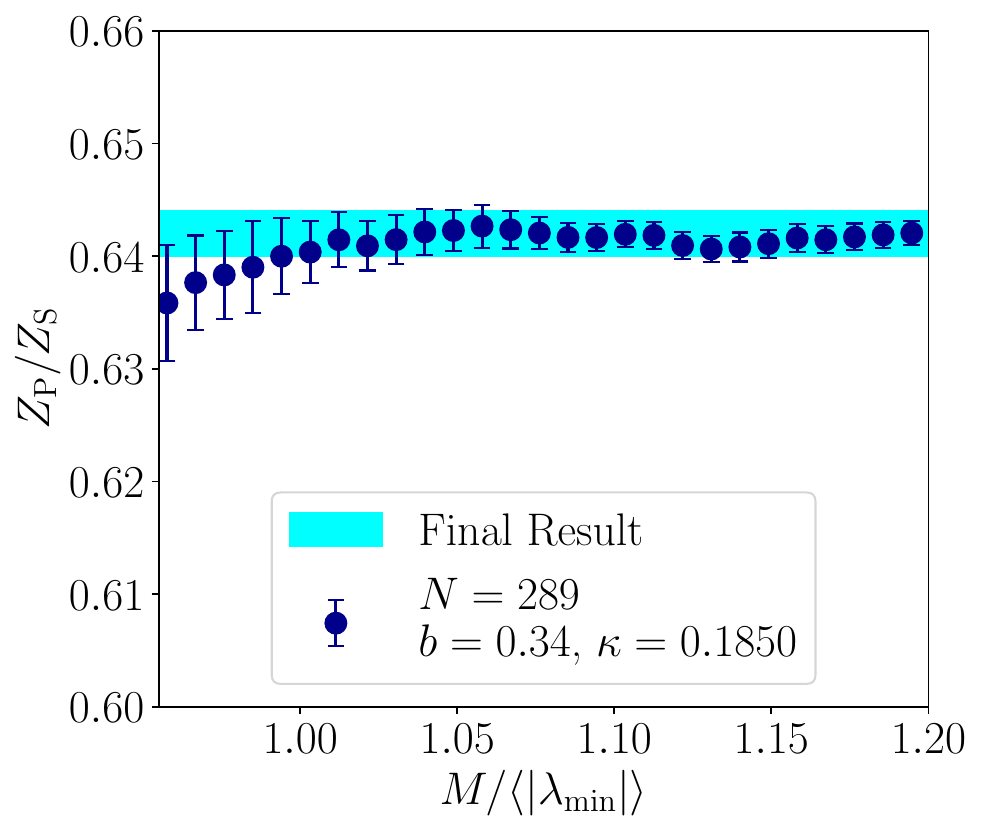}
\caption{Example of calculation of the ratio of non-singlet pseudoscalar and scalar renormalization constants.}
\label{fig:ZPZS}
\end{subfigure}
\caption{Numerical results using spectral methods. Figures taken from~\cite{Bonanno:2024bqg}.}
\end{figure}

\noindent The gluino condensate is obtained from the BC formula via the Giusti--L\"uscher method~\cite{Giusti:2008vb,Luscher:2010ik,Bonanno:2019xhg,Athenodorou:2022aay,Bonanno:2023xkg,Bonanno:2023ypf}.
\begin{itemize}
\item We solved numerically $\left(\gamma_5 D_{_\mathrm{W}}[U]\right) u_{_\lambda} = \lambda \, u_{_\lambda}$ for the first $\mathcal{O}(100)$ eigenvalues.
\item Counting the number of modes below a certain threshold $M$ we obtained the mode number $\braket{\nu(M,m)} = \braket{\# \vert \lambda \vert \le M}$.
\item The gluino condensate is obtained via (here $V=a^4$ with $a$ the lattice spacing):
\beq
\Sigma_\R = \frac{\pi}{4 V} \sqrt{1-\frac{m_\R^2}{M_\R^2}} \left[\frac{\mathrm{d} \braket{\nu_\R(M_\R)}}{\mathrm{d} M_\R}\right] & \longleftarrow \text{slope of $\braket{\nu}$ obtained from}\\\nonumber &\qquad\text{linear fit close to $M=\braket{\vert \lambda_{\min}\vert}\simeq m$}.
\eeq
\item Renormalization: $\braket{\nu_\R} = \braket{\nu}$, $M_\R= M/Z_\P$, $m_\R = m / Z_\S^{\zero} = r_\m m / Z_\S$, $r_\m \equiv Z_\S/Z_\S^{\zero}$, where $Z_\S$ and $Z_\P$ are the non-singlet scalar/pseudo-scalar renormalization constants, while $Z_\S^{\zero}$ is the singlet scalar renormalization constant.\footnote{In our original paper~\cite{Bonanno:2024bqg} we overlooked that $r_\m=1+\mathcal{O}(\lambda^2)$ on the lattice due to explicit chiral symmetry breaking of Wilson fermions. We correct it here, with negligible impact on final results, and no change in the conclusions.}
\end{itemize}

On the lattice we can compute the following quantities:
\begin{itemize}
\item The slope of the mode number as a function of the bare threshold $M$, see Fig.~\ref{fig:fit_mode_number}: $s \equiv \frac{\mathrm{d} \braket{\nu}}{\mathrm{d} M}$.
\item The RGI ratio of non-singlet renormalization constants $Z_\P/Z_\S$ using the following ratio of spectral sums~\cite{Giusti:2008vb,Luscher:2010ik,Bonanno:2019xhg,Athenodorou:2022aay}, see Fig.~\ref{fig:ZPZS}:
\beq
\left(\frac{Z_\P}{Z_\S}\right)^2 = \frac{\braket{s_\P(M)}}{\braket{\nu(M)}} \qquad \qquad
s_\P(M) \equiv \sum_{\vert\lambda\vert,\vert\lambda^{\prime}\vert \le M} \left\vert u_\lambda^\dagger \gamma_5 u_{\lambda^\prime} \right\vert^2.
\eeq
\item Bare subtracted gluino mass:
$
am = 1/(2\kappa) - 1/(2\kappa_{\scriptscriptstyle{\rm crit}}) = a m_\R Z_\S^{\zero} = a m_\R Z_\S / r_\m.
$
\end{itemize}
Combining these quantities we obtain:
\beq
\frac{\Sigma_{\R}}{Z_\S} = \frac{Z_\P}{Z_\S}\frac{\pi}{4V} \sqrt{1-\left(\frac{Z_\P}{Z_\S}\frac{m}{M} r_\m\right)^2} \, s.
\eeq
Since we do not have non-perturbative estimates of the scheme-/scale-dependent constant $Z_\S$ and of the RGI ratio $r_\m = Z_\S / Z_\S^{\zero}$, we relied on 2-loop perturbation theory to compute them in the $\overline{\mathrm{MS}}$ scheme~\cite{Skouroupathis:2007jd}:
\beq
Z_{\S}^{(\MS)}\left(\mu=\frac{1}{a}, \lambda_{\scriptscriptstyle{\rm L}}\right) &=& 1 - \frac{12.9524103(1)}{(4\pi)^2} \lambda_{\scriptscriptstyle{\rm L}} - \frac{60.68(10)}{(4\pi)^4} \lambda_{\scriptscriptstyle{\rm L}}^2  + \mathcal{O}\left(\lambda_{\scriptscriptstyle{\rm L}}^3\right)\\
\nonumber\\
{Z_\S^{\zero}}^{(\MS)}\left(\mu=\frac{1}{a}, \lambda_{\scriptscriptstyle{\rm L}}\right) &=& Z_\S^{(\MS)}\left(\mu=\dfrac{1}{a}, \lambda_{\scriptscriptstyle{\rm L}}\right) - \dfrac{107.76(1)}{(4\pi)^4}\lambda_{\scriptscriptstyle{\rm L}}^2 + \mathcal{O}\left(\lambda_{\scriptscriptstyle{\rm L}}^3\right).
\eeq

In these expressions, in order to accelerate the convergence of perturbation theory, we used the 1-loop perturbative expression of the lattice bare coupling in terms of improved couplings:
\beq
\lambda_{\scriptscriptstyle{\rm L}} = \lambda_{\t}^{(\s)} - 2 b_0 \left(\lambda_{\t}^{(\s)}\right)^2 \log\left(\Lambda_{\s}/\Lambda_{\scriptscriptstyle{\rm L}}\right),
\eeq
with $\lambda_\t^{(\I)} = 1/(bP)$, $\lambda_\t^{(\E)} = 8(1-P)$, $\lambda_\t^{(\Ep)} = -8\,\log(P)$, and $P$ the expectation value of the plaquette. The ratio of $\Lambda$-parameters is given by~\cite{Weisz:1980pu,Perez:2017jyq}:
\beq
\begin{aligned}
\dfrac{\Lambda_{\scriptscriptstyle{\rm L}}}{\Lambda_{\I} } &= \dfrac{\Lambda_{\scriptscriptstyle{\rm L}}}{\Lambda_{\MS}} \times 2.7373 , \\
\dfrac{\Lambda_{\scriptscriptstyle{\rm L}}}{\Lambda_{\E} } &= \dfrac{\Lambda_{\scriptscriptstyle{\rm L}}}{\Lambda_{\MS}} \times 29.005,  \\
\dfrac{\Lambda_{\scriptscriptstyle{\rm L}}}{\Lambda_{\Ep} } &= \dfrac{\Lambda_{\scriptscriptstyle{\rm L}}}{\Lambda_{\MS}} \times 5.600 ,\\
\dfrac{\Lambda_{\MS}}{\Lambda_{\scriptscriptstyle{\rm L}}} &= 73.467 .
\end{aligned}
\eeq

\subsection{The gluino condensate from the GMOR equation}

Using standard techniques, in~\cite{Butti:2022sgy} we computed the pion mass $m_\pi$ from the exponential time decay of the temporal pion-pion correlator, the pion decay constant,
\beq
\frac{F_\pi}{N Z_\A} = \frac{1}{\sqrt{2} N m_\pi} \langle 0 \vert A_4(x=0) \vert \pi(\vec{p}=0) \rangle,
\eeq
and the PCAC (Partially Conserved Axial Current) gluino mass, related to the renormalized one by:
\beq
m_{\PCAC} = \frac{Z_\P}{Z_\A} m_\R.
\eeq
Combining these quantities, and the ones computed before, one has:
\beq\label{eq:Sigma_GMOR}
\frac{F_\pi}{N} = \frac{F_\pi}{N Z_\A} \frac{m}{m_{\PCAC}} \frac{Z_\P}{Z_\S} r_\m, \\
\nonumber\\
\qquad \frac{\Sigma_{\R}}{Z_\S} = \frac{1}{2} F_\pi^2 \frac{m_\pi^2}{m} \frac{1}{r_\m} \,.
\eeq

\vspace{\baselineskip}
\subsection{The $\Lambda$-parameter in the NSVZ scheme}

\begin{wrapfigure}[11]{c}{0.42\textwidth}
\centering
\vspace*{-0.5\baselineskip}
\includegraphics[width=0.42\textwidth]{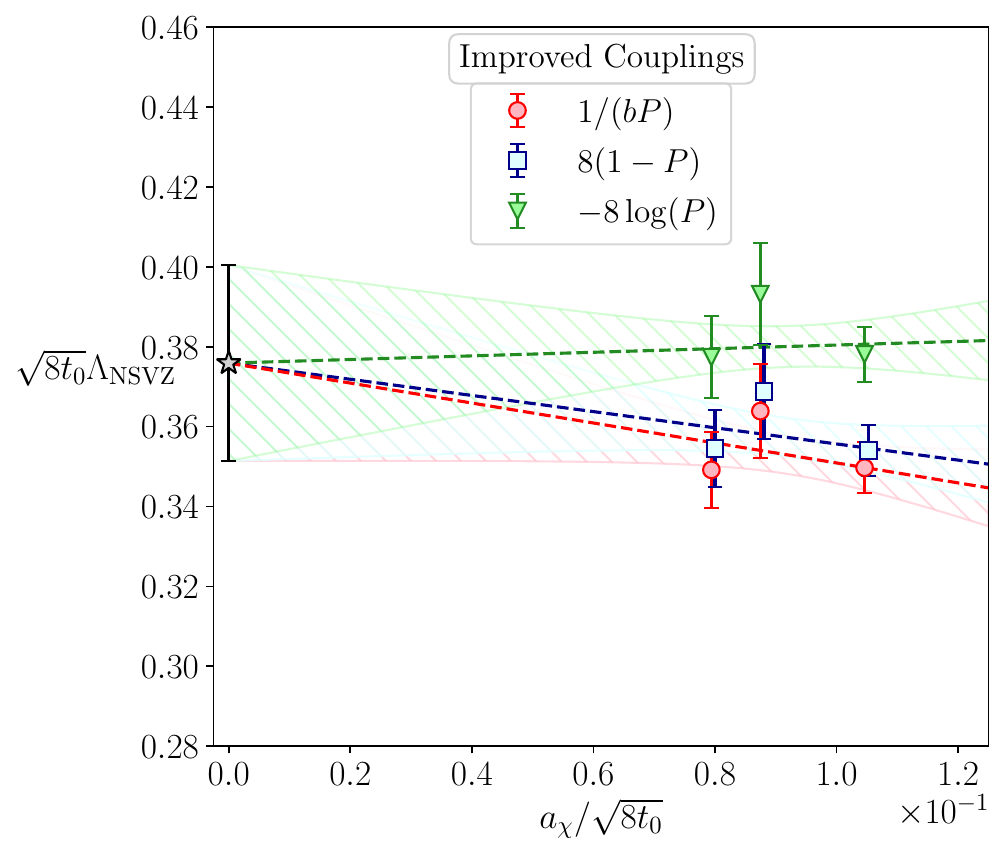}
\caption{Extrapolation of $\Lambda_{\NSVZ}$ towards the SUSY limit. Figure taken from~\cite{Bonanno:2024bqg}.}
\label{fig:Lambda_NSVZ_SUSY}
\end{wrapfigure}

We computed the $\Lambda$-parameter using 2-loop asymptotic scaling, and the 3 improved couplings introduced earlier (here $a_{\scriptscriptstyle{\chi}}$ denotes the lattice spacing extrapolated towards the massless gluino limit):
\beq
\begin{aligned}
\sqrt{8t_0}\Lambda_{\NSVZ} &= \lim_{a_\chi\to 0} \frac{\Lambda_\NSVZ}{\Lambda_\MS} \frac{\Lambda_{\MS}}{\Lambda_{\s}}\frac{\sqrt{8t_0}}{a_{\chi}} \exp\left\{-f\left(\lambda_\t^{(\s)}\right)\right\},\\
f(x) &= \frac{1}{2 b_0}\left[\frac{1}{x} + \frac{b_1}{b_0} \log(b_0 x)\right].
\end{aligned}
\eeq
where $\Lambda_\NSVZ/\Lambda_\MS = \mathrm{e}^{-1/18}$~\cite{Finnell:1995dr}.

The 3 improved couplings yield results in very good agreement, as shown in Fig.~\ref{fig:Lambda_NSVZ_SUSY}. We quote the following final result, obtained imposing a common continuum limit for the 3 different determinations:
\beq\label{eq:final_res_Lambda}
\begin{aligned}
\sqrt{8t_0}\Lambda_{\NSVZ} &= 0.376(25), \qquad \qquad \sqrt{8t_0}\Lambda_{\MS} &= 0.397(26).
\end{aligned}
\eeq

\subsection{Converting the renormalized condensate into the RGI one}

In the previous subsections we described how we computed the scheme-/scale-dependent renormalized chiral condensate $\Sigma_\R^{(\s)}(\mu)$ in the $\overline{\mathrm{MS}}$ scheme at the lattice scale $\mu = 1/a$, with $a$ the lattice spacing. In order to compare our results with the analytic ones, we need to convert it into an RGI quantity.

To this end, we rely again on 2-loop perturbation theory as follows:
\beq
\Sigma_{\RGI} = {\cal A} \, 2 b_0 \lambda_{\t}^{(\MS)}(a\mu=1) \, \left[ 1  +  \frac{d_1^{(\MS)}\!-\!2b_1}{2b_0} \lambda_{\t}^{(\MS)}(a\mu=1) \right] \Sigma_\R^{(\MS)}(a\mu=1), \quad \mathcal{A}=8\pi^2/(9N^2),
\eeq
with $d_1^{(\MS)} = 32/(4\pi)^4$~\cite{Vermaseren:1997fq}, and where the renormalized coupling in the $\overline{\mathrm{MS}}$ scheme at the scale $\mu = 1/a$ was computed in 2-loop perturbation theory as follows:
\beq
2 b_0\lambda_{\t}^{(\MS)}(a\mu=1) = -\frac{1}{\log(a \Lambda_{\MS})} - \frac{b_1}{2 b_0^2} \frac{ \log\left[-2\log(a \Lambda_{\MS})\right]}{\log^2(a \Lambda_{\MS})} \, .
\eeq
The product $a \Lambda_\MS$ was practically computed as $a \Lambda_\MS= \left(a/\sqrt{8t_0}\right) \times \sqrt{8t_0} \Lambda_\MS$, with $\sqrt{8t_0}\Lambda_\MS$ the quantity in Eq.~\eqref{eq:final_res_Lambda}.

\subsection{The $N$-dependence of the gluino condensate}

\begin{figure}[!htb]
\centering
\includegraphics[scale=0.29]{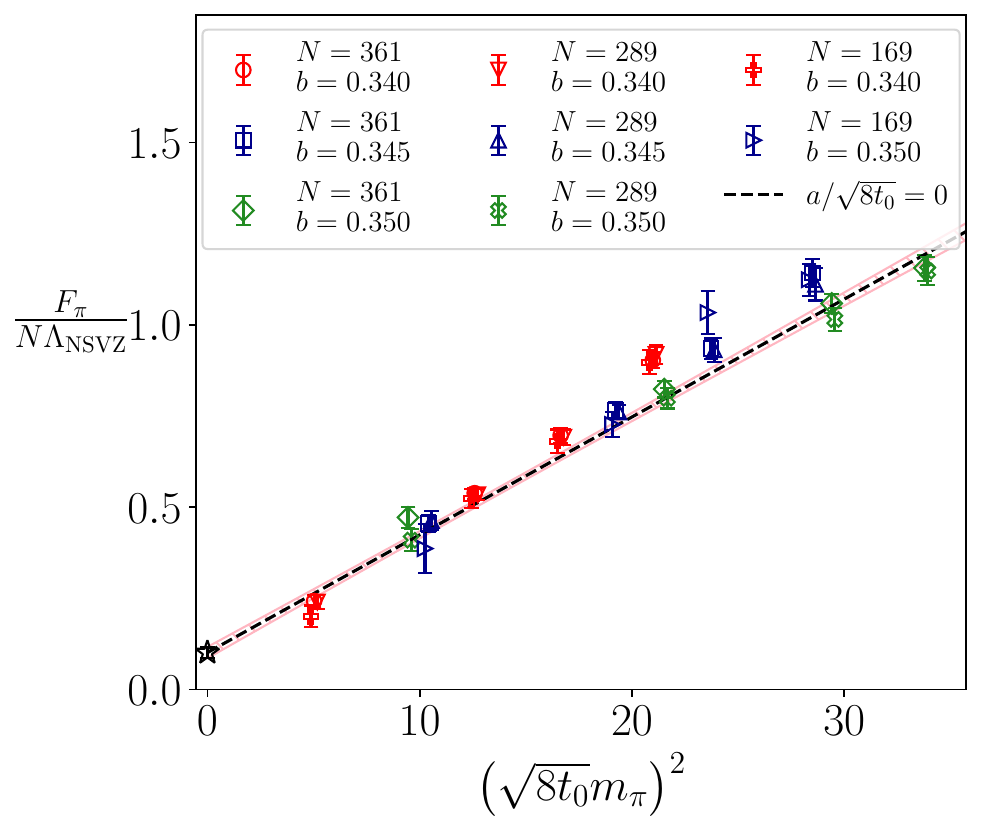}
\includegraphics[scale=0.29]{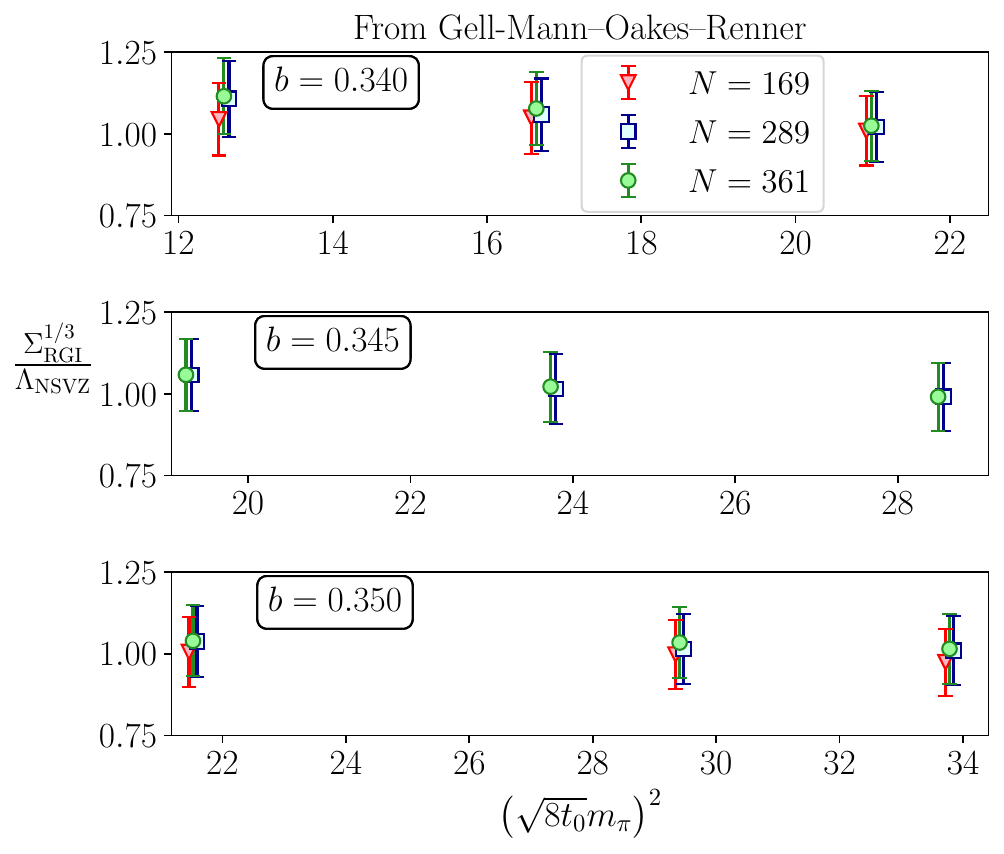}
\includegraphics[scale=0.29]{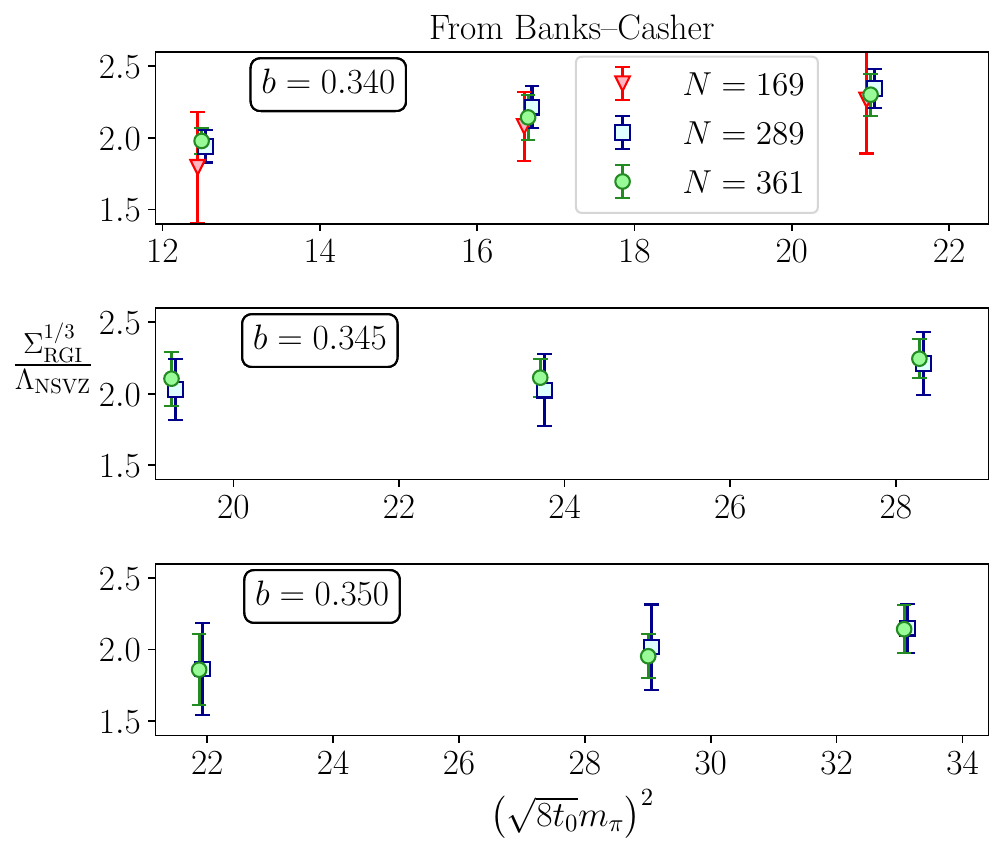}
\caption{Left panel: calculation of $F_\pi/N$ in the SUSY limit. Central and right panels: collection of our results for the third root of the RGI condensate in units of the SUSY scale in the NSVZ scheme $\Lambda_\NSVZ$ obtained with the GMOR and the BC formulas respectively. Figures adapted from~\cite{Bonanno:2024bqg}.}
\label{fig:Ndep_Sigma_RGI}
\end{figure}

We are now ready to compute the RGI gluino condensate from the BC and the GMOR relations in units of $\Lambda_\NSVZ^3$. Note that Eq.~\eqref{eq:Sigma_GMOR} was computed using the value of $F_\pi$ extrapolated towards the SUSY limit (via a joint chiral/continuum extrapolation), cf.~Fig.~\ref{fig:Ndep_Sigma_RGI} (left panel):
\beq
\frac{F_\pi}{N\Lambda_\NSVZ} = 0.101(15) \,.
\eeq
All our determinations of $\Sigma_\RGI$ are shown in Fig.~\ref{fig:Ndep_Sigma_RGI} (central/right panels). Given the displayed $N$-dependence, our numerical results rule out all but the Weak Coupling (WC) calculation, cf.~Eq.~\eqref{eq:Sigma_Inst}.

\section{Conclusions: the RGI gluino condensate in the SUSY limit}

\begin{wrapfigure}[13]{c}{0.42\textwidth}
\centering
\includegraphics[width=0.42\textwidth]{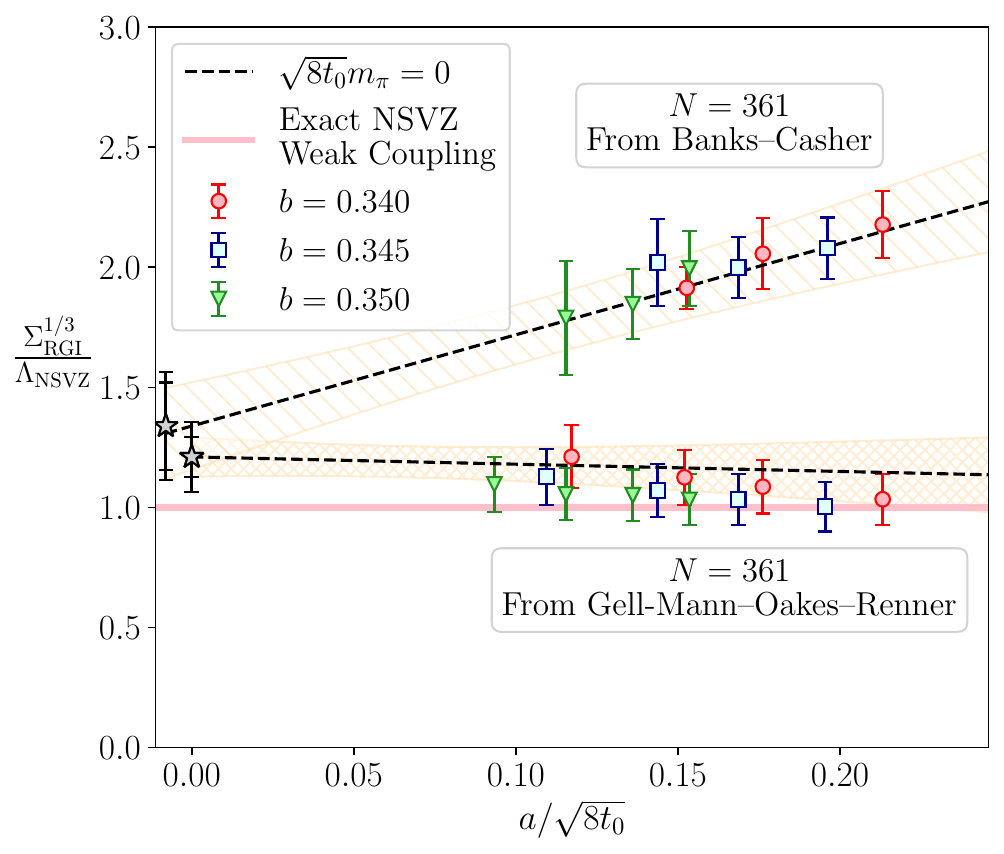}
\caption{Extrapolation towards the SUSY limit of the RGI gluino condensate determined from the BC and the GMOR formulas for the largest value of $N$ explored. Figure adapted from~\cite{Bonanno:2024bqg}.}
\label{fig:final}
\end{wrapfigure}

We extrapolate our results for $\Sigma_\RGI^{1/3}/\Lambda_\NSVZ$ towards the SUSY limit performing a joint chiral-continuum extrapolation as follows:
\beq
\left(\frac{\Sigma_\RGI^{1/3}}{\Lambda_\NSVZ}\right)(a,m_\pi) = \frac{\Sigma_\RGI^{1/3}}{\Lambda_\NSVZ} + c_1 \frac{a}{\sqrt{8t_0}}+c_2 \left(8t_0m_\pi^2\right).
\eeq
The SUSY-limit extrapolations, shown in Fig.~\ref{fig:final}, yield:
\beq
\begin{aligned}
\frac{\Sigma_\RGI}{\Lambda_{\NSVZ}^3} &= [1.34(18)_{\rm stat}(13)_{\rm syst}]^3 \\
&= 2.39(97)_{\rm stat}(72)_{\rm syst}, \qquad \qquad \text{(BC)} \, .
\end{aligned}
\eeq
\beq
\begin{aligned}
\frac{\Sigma_\RGI}{\Lambda_{\NSVZ}^3} &= [1.21(08)_{\rm stat}(12)_{\rm syst}]^3 \\
&= 1.77(35)_{\rm stat}(53)_{\rm syst}, \qquad \qquad \text{(GMOR)} \, .
\end{aligned}
\eeq
In the SUSY limit the two determinations from the GMOR and BC relations give agreeing results. Both are compatible with the WC instanton calculation:
\beq
\frac{\Sigma_\RGI}{\Lambda_{\NSVZ}^3}= 1, \qquad \qquad \text{(exact NSVZ analytic WC result)} \, .
\eeq
We quote the GMOR determination as our final lattice result, as this is the most precise one:
\beq
\frac{\Sigma_\RGI}{\Lambda_{\NSVZ}^3} = 1.77(35)_{\rm stat}(53)_{\rm syst} = 1.77(65), \qquad \qquad \text{(final lattice result)} \, .
\eeq
We stress that we added a conservative 30\% systematic error to our final extrapolations to take into account the perturbative renormalization we employed. This is the dominant source of uncertainty.

\section*{Acknowledgments}
It is a pleasure to thank José L.~F.~Barbón, Georg Bergner, Gregorio~Herdo\'iza, Nikolai Husung, Haralambos Panagopoulos, Carlos Pena and Erich Poppitz for useful discussions. This work is partially supported by the Spanish Research Agency (Agencia Estatal de Investigaci\'on) through the grant IFT Centro de Excelencia Severo Ochoa CEX2020-001007-S, funded by MCIN/AEI/10.13039/501100011033, and by grant PID2021-127526NB-I00, funded by MCIN/ AEI/10.13039/501100011033 and by “ERDF A way of making Europe”. We also acknowledge support from the project H2020-MSCAITN-2018-813942 (EuroPLEx) and the EU Horizon 2020 research and innovation programme, STRONG-2020 project, under grant agreement No 824093. P.~B.~is supported by Grant PGC2022-126078NB-C21 funded by MCIN/AEI/10.13039/ 501100011033 and “ERDF A way of making Europe”. P.~B.~also acknowledges support by Grant DGA-FSE grant 2020-E21-17R Aragon Government and the European Union - NextGenerationEU Recovery and Resilience Program on ‘Astrofísica y Física de Altas Energías’ CEFCA-CAPA-ITAINNOVA. K.-I.~I.~is supported in part by MEXT as "Feasibility studies for the next-generation computing infrastructure". M.~O.~is supported by JSPS KAKENHI Grant Number 21K03576. Numerical calculations have been performed on the \texttt{Finisterrae~III} cluster at CESGA (Centro de Supercomputaci\'on de Galicia). We have also used computational resources of Oakbridge-CX at the University of Tokyo through the HPCI System Research Project (Project ID: hp230021 and hp220011).


\providecommand{\href}[2]{#2}\begingroup\raggedright\endgroup


\end{document}